\documentclass[11pt]{article}

\usepackage{paper}

\crefname{figure}{Figure}{Figures}

\title{\textbf{From Defense to Advocacy: Empowering Users to\\
  Leverage the Blind Spot of AI Inference}}

\author[1]{Yumou Wei}
\author[2]{John Carney}
\author[1]{John Stamper}
\author[2]{Nancy Belmont}
\affil[1]{Carnegie Mellon University, Pittsburgh, PA, USA\\
  \fontshape{n}\selectfont\{yumouw, jstamper\}@andrew.cmu.edu}
\affil[2]{MARi LLC, Alexandria, VA, USA\\
  \fontshape{n}\selectfont \{john.carney, nancy.belmont\}@mari.com}

\runningtitle{From Defense to Advocacy}

\hypersetup{
  pdftitle={From Defense to Advocacy: Empowering Users to Leverage the Blind Spot of AI Inference},
  pdfauthor={Yumou Wei, John Carney, John Stamper, Nancy Belmont},
  pdfkeywords={Data Privacy, Advocacy Agents}}

\begin{document}

\maketitle

\begin{abstract}
Most privacy regulations function as a passive defensive shield that users must wield themselves. Users are incessantly asked to ``opt-in'' or ``opt-out'' of data collection, forced to make defensive decisions whose consequences are increasingly difficult to predict. Viewed through the Johari Window, a psychological framework of self-awareness based on what is known and unknown to self and others, current policies require users to manage information in the Arena and shield the Façade through notice and consent. However, as organizations increasingly use AI to make inferences, the rapid expansion of Blind Spot---attributes known to algorithms but unknown to the user---emerges as a critical challenge. We illustrate how current regulations fall short and leave this Blind Spot unguarded because they focus on data collection rather than inference. Considering this challenge, we propose a paradigm shift from defensive privacy management to proactive privacy advocacy. We argue for the necessity of Privacy Advocacy Agents capable of simulating possible inferences under social norms to harness the power of AI inference. By illuminating the hidden inferences that users can strategically leverage or suppress, these agents not only restrain the growth of the Blind Spot but also transform the unknown information into personal assets for users. This way, we can foster a flow of personal information that is equitable, transparent, and individually beneficial in the age of AI.
\end{abstract}

% ---------------------------------------------------------------------
\section{Introduction}
\label{sec:intro}

Every day, almost every internet user faces a vexing conundrum: accepting all cookies, refusing unnecessary cookies, or refusing all cookies? These ubiquitous cookie banners are a microcosm of the ``opt-in'' or ``opt-out'' mechanisms embedded in many contemporary privacy regulations, such as the General Data Protection Regulation (GDPR) in the European Union and the California Consumer Privacy Act (CCPA) in the United States. These regulations generally require organizations to \emph{notify} users about how data is collected and obtain users' \emph{consent} before collecting or processing personal data. This \textbf{notice-and-consent} paradigm implicitly assumes that users can protect their privacy by simply choosing to accept or refuse data collection and fully understand the consequences of each choice---as though raising a defensive shield against an incoming threat. Consequently, it is primarily the users' responsibility to manage privacy risks, even if they lack the time, expertise, or background knowledge required to make informed decisions about whether to raise or lower this shield~\citep{Cate2013}.

With the accelerating development of artificial intelligence (AI), this traditional privacy protection paradigm is facing unprecedented challenges. Not only do organizations collect personal data, but they also increasingly use sophisticated AI algorithms to infer \emph{additional} information about a user, often beyond what the user would have expected. A frequently cited example is Target sending maternity product coupons to a female customer based on her purchase history before her family was even aware of her pregnancy~\citep{duhigg2012how}. Although this example predates modern advanced AI, it illuminates a more general phenomenon that users are often unaware of the \emph{hidden inferences} that can be made about them from seemingly trivial data.

We argue that this phenomenon can be better understood through the Johari Window model~\citep{luft1955johari}. This model was originally proposed to understand self-awareness. It divides what is known and unknown about oneself and others into four quadrants (\cref{fig:johari}): \textbf{Arena} (known to both), \textbf{Façade} (only known to self), \textbf{Blind Spot} (only known to others), and \textbf{Unknown} (known to neither). We repurpose this model to examine privacy management as orchestrating the information flows between these four quadrants. Viewed through this model, the current notice-and-consent paradigm focuses on the information flow between Arena and Façade, requiring users to decide what to disclose to others (moving information into the Arena) and what to keep secret (hiding information behind the Façade). For Blind Spot and Unknown, current regulations do not provide sufficient protection, even though it is only a matter of time before AI inference extracts and moves information from Unknown to Blind Spot. AI inference is rapidly growing and driving a rapid expansion of the Blind Spot as organizations extract additional knowledge about the users from legally collected data beyond the users' awareness. Users are often in a passive and vulnerable position, unable to anticipate, observe, or contest many of the hidden inferences made about them. Given the unstoppable growth of AI inference across many fields, how can a user effectively protect their privacy in these unique times?

In this paper, we explore a paradigm shift from defensive privacy management to \textbf{proactive privacy advocacy}. Through a critical analysis of existing privacy regulations, we show that the traditional notice-and-consent paradigm only covers data collection and is deficient for scenarios in which AI uses legally collected and sourced data to infer additional personal attributes that are not covered by the consent. To combat unsolicited AI inference, we argue that new systems must be built to advocate for users' active participation in managing and utilizing personal information. To this end, we conceptualize \textbf{Privacy Advocacy Agents (PAAs)}, which help users identify and understand the Blind Spot created by possible AI inference made about the users and devise strategies to utilize or suppress this inference before it is exploited by others, transforming privacy management from passive defense to proactive advocacy.

% ---------------------------------------------------------------------
\section{The Deficiencies of Binary Defense}
\label{sec:deficiencies}

Most current privacy regulations, especially those based on the notice-and-consent principle, rely on a fundamental ``binary defense'' mechanism to protect privacy. Under this mechanism, users are effectively forced to choose between two extremes: either accept data collection and gain access to the service or refuse data collection and typically lose access. These binary choices reduce the complex and dynamic trust negotiation process~\citep{Winsborough2000} to a one-time decision. Although this mechanism provided a practical heuristic in the early days of data processing dominated by simple aggregation, storage, and rule-based analysis, this simplified defense mechanism is increasingly inadequate in contemporary data processing environments, especially given the proliferation of AI systems that use sophisticated algorithms for accurate inference~\citep{Solove2025Artificial}.

We conducted a critical analysis of the existing privacy regulations, demonstrating that this binary defense mechanism is structurally deficient in addressing the new privacy challenges that AI inference creates. These deficiencies are not merely a matter of legal loopholes or poor enforcement, but a deeper design flaw recurring at various governance levels, including governments, organizations, and institutions. It is important to clarify, though, that we are not arguing that the current privacy laws and standards completely neglect inference, profiling, or subsequent use of personal data. Rather, our argument is based on the design structure of these laws and standards. These frameworks are primarily designed to protect discrete data records, identifiable collection activities, and observable disclosures. Therefore, they are weakly regulated in terms of the generation, dissemination, and reuse of probabilistic inference, especially when such inference exceeds clearly defined sensitive categories or legally recognized decision-making scope. Therefore, our analysis should not be interpreted as a claim of legal deficiencies but rather as a discussion of the systemic misalignment between regulatory forms and inference practices.

\subsection{The Governmental Paradigm: The Limits of Privacy Self-Management}

At the government level, major privacy regulations, such as the GDPR and CCPA, remain heavily influenced by the ``privacy self-management'' model~\citep{Solove2013}. This model assumes that users can effectively manage their own privacy if they have sufficient information and choice. In practice, the law often assumes that transparency plus choice is sufficient to protect privacy. Therefore, under this model, regulatory compliance primarily focuses on the moment of data collection. If the user has been informed and has given their ``consent'', then data collection and subsequent processing are considered legal and legitimate.

However, this model overlooks several key issues. First, privacy self-management hinges on the limited cognitive and information processing abilities~\citep{Simon1982} of the users. An average internet user needs to read and understand a plethora of lengthy and complex privacy policies filled with legal jargon, and making truly informed decisions is often impossible~\citep{Acquisti2005,Cate2013,McDonald2008}. These privacy notices often face a dilemma~\citep{Solove2013}: to be legally comprehensive, privacy statements must be exhaustive and not obscure legally significant details; however, to be accessible to average users, they must be concise and clear, not overwhelming users with esoteric legal terms.

Moreover, this model tends to ignore the dynamic aspect of data processing. It often assumes that data is all that has been collected---no more or no less---and that data can be stably categorized. Certain categories of data, such as Social Security numbers or medical records, are categorized as inherently sensitive, whereas other categories, such as mouse clicks or keystrokes, are considered harmless at the time of collection. Thus, privacy protection is constructed around designated ``sensitive input'' boundaries. However, in AI-powered systems, data is not merely what has been collected, as it can also be used to generate new insights that may far exceed what the users expect. Some seemingly harmless, discrete data points, such as typing rhythm, word selection, and geo-location patterns, can be aggregated and analyzed to reveal highly sensitive personal information, analogous to what legal scholars refer to as the ``mosaic theory''~\citep{Kerr2012}. Organizations can legally and compliantly collect ``non-sensitive'' personal data without triggering the defense mechanism and then use the data to perform complex inference to draw conclusions that the users did not authorize or foresee, such as health status or financial risks. In this case, user privacy is effectively weakened because users can neither control nor understand these inference processes and the outcomes. Although the GDPR partially regulates user profiling and automated decision-making (Articles 22 and 13-15 in particular), these provisions are limited in scope, complex in procedure, and difficult to implement in practice~\citep{Wachter2018Counterfactual,Wachter2017Why}. Therefore, the law often legitimizes data collection while only weakly regulating the subsequent inference processes.

\subsection{The Organizational Paradigm: Security as a Proxy for Privacy}

At the organizational level, many organizations view privacy protection as a compliance obligation rather than a substantive question of user agency. Privacy policies and practices often focus on complying with legal requirements rather than understanding and respecting users' privacy preferences. This compliance-oriented privacy management model further reinforces the binary defense mechanism by emphasizing obtaining user consent at the time of data collection while equivocating on subsequent stages of data processing, where AI inference may happen. Organizations typically assess the privacy risks of their data-processing activities through Privacy Impact Assessments (PIAs), but these assessments often focus on the collection and storage phases rather than the inference and analysis phases. Mainstream organizational-level privacy management frameworks, such as the NIST privacy framework and the ISO/IEC 27701:2025 standards, typically emphasize protecting data privacy through technological and organizational measures. However, these frameworks often focus on preventing unauthorized access and data breaches (a security-oriented approach), rather than managing the privacy risks associated with AI inference. Consequently, organizations often overlook the privacy challenges posed by AI inference, continuing to rely on the binary defense mechanism to remain compliant.

In an AI-driven organization, data analysts, product managers, and even AI models are often granted legitimate access to raw user data for optimizing, personalizing, or improving services. Although organizations may implement stringent data protection protocols to prevent unauthorized access, such as encryption and access controls, these measures do not prevent authorized users or models from using the data for inference. Once such access is authorized---that is, once the binary switch is turned to ``allowed''---the security mechanism has essentially done its job, and it remains unfettered what new knowledge is generated from this access.

For example, consider an e-commerce platform that implements strong security measures to protect users' purchase records from unauthorized access. This security framework can prevent hackers from stealing users' purchase records; however, by design, it cannot prevent \emph{authorized} models from using the same purchase records to infer sensitive information about a user. The outcomes of such inference could be used for personalized advertising, credit scoring, or other decisions without the user's knowledge. In this sense, organizational defenses typically protect data as an asset, while providing relatively weak protection for \emph{data subjects} as a person. The risks associated with the former are managed through property-like access and control measures; risk management of the latter, if ever implemented, is achieved indirectly through the derivative effects of the security measures. This asymmetry becomes particularly pronounced when the results of inference (rather than the original data) become the primary driver of decision-making.

\subsection{The Institutional Paradigm: The Accuracy Trap}

At the institutional level, privacy regulations typically focus on ensuring the accuracy and integrity of data, especially in areas involving credit reporting, employment screening, and medical records. For example, the Fair Credit Reporting Act (FCRA) in the United States requires credit reporting agencies to ensure the accuracy of their reports and grants consumers the right to correct misinformation~\citep{FCRA2023}. However, this focus on accuracy often overlooks the privacy risks associated with AI inference, particularly when the inference is based on accurate data but draws conclusions that could negatively impact a person. For instance, an AI system might infer a person's health conditions based on accurate purchase records, and these inferences could fuel the decisions to deny a loan or increase insurance rates. Although these inferences may be based on accurate data, they still infringe on an individual's privacy rights when unsolicited. Furthermore, institutional privacy regulations often lack the oversight of the AI inference processes, leaving no effective appeal or correction mechanisms for users facing unfair decisions.

Individuals have the right to correct facts, but not the right to challenge the interpretations derived from the facts---we call this the ``accuracy trap''. This trap is particularly pronounced in the context of AI inference, as AI systems often make complex inferences that far exceed a user's expectations. Even if the inference is based on accurate data, users may not be able to challenge its illegality or unfairness because the law does not grant them such a right. Furthermore, the accuracy trap can lead to a form of hidden harm where users are unaware of the negative impact that the inference has on their social standing. The possibilities of such a hidden harm place users in a passive and vulnerable position because they are unable to identify or address these invisible threats, not to mention raising their defensive shields.

\subsection{A Johari Window Analysis of the Structural Deficiencies}

\begin{figure}[t]
  \centering
  \includegraphics[width=0.55\linewidth]{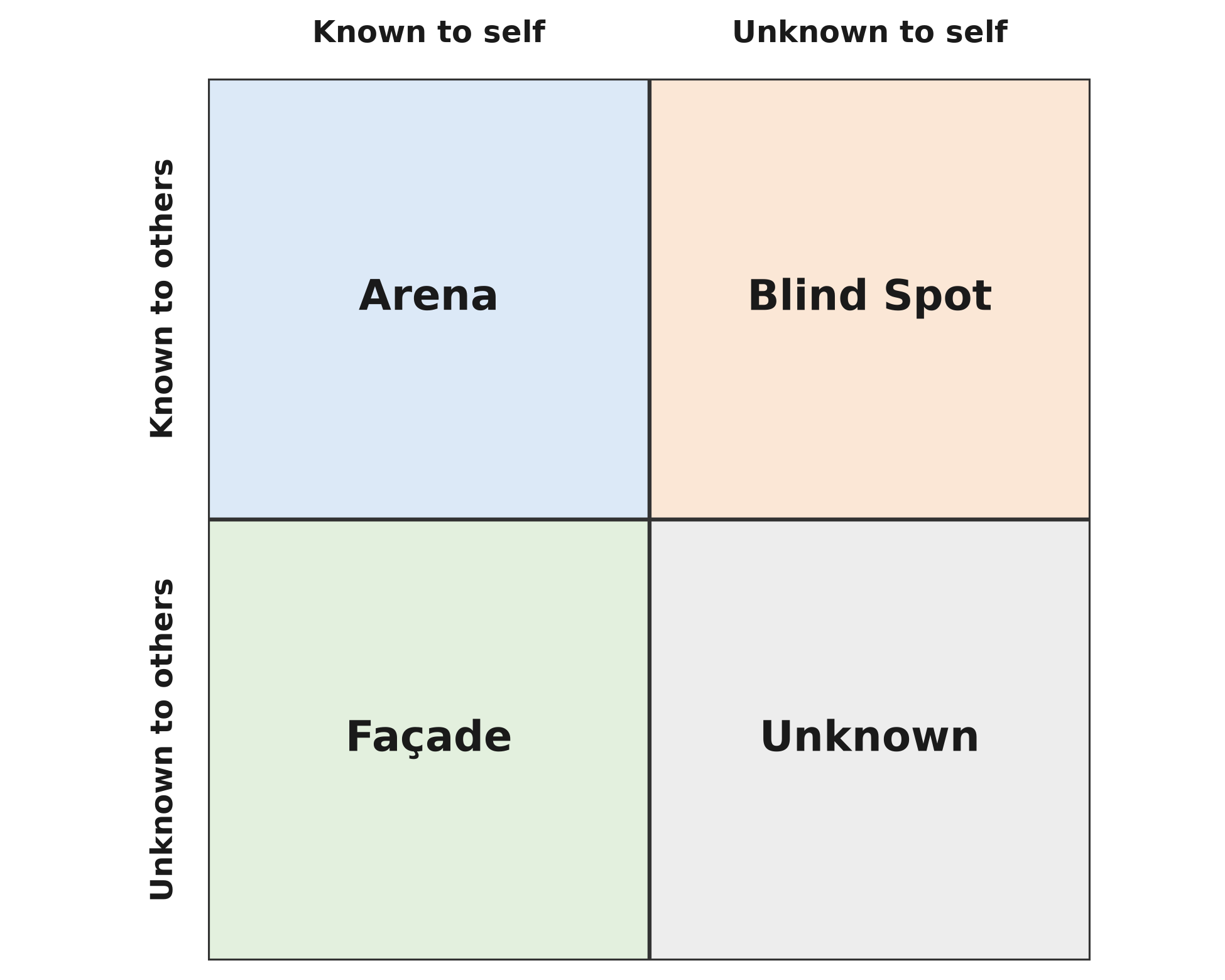}
  \caption{A Johari Window divides self-awareness into four quadrants based on what is known and unknown.}
  \label{fig:johari}
\end{figure}

To better understand the shortcomings of the current privacy regulations in addressing the privacy challenges posed by AI inference, we apply the Johari Window model~\citep{luft1955johari} to analyze the structural deficiencies of binary defense. As a psychological heuristic originally used to map interpersonal consciousness, the Johari Window (\cref{fig:johari}) divides self-awareness into four quadrants based on what is known and unknown to self and others: \textbf{Arena} (known to both), \textbf{Façade} (only known to self), \textbf{Blind Spot} (only known to others), and \textbf{Unknown} (known to neither). These four quadrants can help us understand the different personal information flowing between these categories and reveal the limitations of the current privacy protections.

\textbf{Arena} includes information that a person is willing to share with others, such as name, email address, and social media profile. Current privacy regulations primarily focus on managing information in the Arena, requiring organizations to adhere to the principles of transparency when collecting and using information that belongs to this quadrant. Through notice and consent, users can control what information is disclosed and how it is used. However, as AI systems increasingly use public data in the Arena to make novel inferences, the flow of information in this quadrant has become more intricate. While users may consent to sharing certain information, they often cannot foresee how this information will be used to generate new insights, leading to a failure of privacy protections.

\textbf{Façade} contains information that users wish to keep confidential, such as medical records, financial information, and private communications. Privacy regulations play a crucial role in restricting access to and use of sensitive information in this quadrant. For example, HIPAA regulations specifically protect the privacy of medical information, requiring healthcare institutions to take strict measures to prevent unauthorized access and disclosure. However, as we discussed earlier, AI systems can still paint a detailed, mosaic picture of what is behind a user's Façade by inferring sensitive information from non-sensitive data, circumventing privacy protections and creating a false sense of security.

\textbf{Blind Spot} contains a user's attributes that have been discovered by algorithms but remain unknown to the user. The user is unaware of the automatic discovery or even the existence of these attributes. For example, an AI system can calculate addiction risks or bankruptcy likelihood, which an average user cannot easily quantify. As AI's inference capabilities increase, a user's Blind Spot is rapidly expanding and becoming a growing privacy challenge. Current privacy regulations often neglect this quadrant because, as we argued, they primarily regulate data collection rather than inference. Since users are not even aware of their Blind Spot, they cannot effectively consent or object to extracting new information from this precious source, nor can they conduct privacy assessments of the risks that they are unaware of. This is not merely a usability issue; it is a structural problem under the notice-and-consent mechanism.

\textbf{Unknown} encompasses attributes that are unknown to both the user and the algorithm. This information may not have been collected or inferred, or it may be difficult to identify because of data scarcity or complexity. The Unknown is not directly addressed in current privacy regulations, but it is akin to an unsharpened sword where either edge can be honed. On the one hand, as data collection and AI inference capabilities continue to advance, what is in Unknown can gradually move into Blind Spot, exposing users to new privacy risks that they are not even aware of. On the other hand, if users can gain insight into their Unknown (before third-party algorithms eventually do), they can potentially move the information into Arena or Façade, thereby utilizing the newly discovered information to their advantage.

A significant threat to privacy in the age of AI is the rapid information movement from Unknown to Blind Spot. This movement means that AI systems have access to more knowledge about a user than the user themselves does, leading to severe information asymmetry and power imbalances. As AI models continue to evolve, they are becoming even more capable of extracting latent attributes from everyday usage data; meanwhile, users retain the illusion of well-protected privacy because no obvious sensitive data is disclosed or collected, as far as the users are made aware. However, through inference, AI systems can already extract sensitive information unexpectedly. This hidden inference challenges the theory of Contextual Integrity~\citep{Nissenbaum2004,Nissenbaum2009privacy}, which defines privacy as the appropriate flow of information relative to contextual norms, because the ``context'' is no longer what a user believes it to be.

In the case of AI inference, the appropriateness of information flow depends not only on the collection and use of data but also on the inference processes and results. Contextual Integrity presupposes that users can recognize their own context. With the Blind Spot expanding, users can no longer accurately assess whether a certain flow of information is appropriate because their awareness of their own information has been disrupted by AI systems. What a user thinks is a flow of information from Unknown to Façade (by not sharing a recently discovered attribute) may, in fact, be a flow from Blind Spot to Façade, if the AI system had already inferred this attribute before the user did. Users may find themselves in a predicament where they cannot control or understand how AI systems are using their data for inference. The users may believe they are conducting a retail transaction, but the AI system may already be inferring their health status or political leanings behind the scenes. This cognitive gap prevents users from effectively fulfilling their privacy self-management responsibilities for Contextual Integrity, and without mechanisms to reveal what might be in the Blind Spot, individuals cannot effectively enforce contextual norms. \textbf{This requires a paradigm shift from defensive tools that block the \emph{input} towards advocacy tools that reveal, question, and negotiate the \emph{output}.} We need to re-examine the privacy-protection paradigms, moving beyond traditional defensive mechanisms and exploring how to proactively manage both the risks and opportunities brought by AI inference.

% ---------------------------------------------------------------------
\section{The Privacy Advocacy Agents}
\label{sec:paa}

The preceding analysis of the binary-defense paradigm shows that the current privacy-protection mechanisms are inadequate in addressing the emergent privacy challenges posed by AI inference. In this section, we describe a new privacy management paradigm representing a shift from binary defense to what we call ``privacy advocacy''. Instead of describing a mature privacy-protection system, we are proposing a novel conceptual framework for reconstructing privacy governance to \emph{empower users} with greater agency. The core assumption is that an increasing number of organizations will use AI inference to expand and exploit everyone's Blind Spot---and this is unstoppable. Simply restricting the flow of information between Arena and Façade (as in the notice-and-consent paradigm) is no longer sufficient to protect privacy, because AI inference largely operates between the Unknown and Blind Spot. We must also enable users to manage the flow of information there or even to utilize that information for their own benefit.

To illustrate this shift, we introduce a novel concept called ``Privacy Advocacy Agent'' (PAA). PAAs refer to a class of systems designed to minimize the epistemic asymmetry between individuals and organizations by revealing and interpreting the Blind Spot that could have been generated by the organization's AI inference, making the inference processes more transparent, understandable, and even helpful to the users. After the PAA helps foreshadow the blind spots by running inference privately for a user, it can also devise strategies for the user to utilize or suppress the extracted insights, transforming privacy management from passive defense to active advocacy.

This framework is distinct from other seemingly similar approaches. In contrast to the transparency tools that aim to explain how the system works, PAAs focus on enabling users to think about \emph{how they are understood and modeled by the system}, rather than simply understanding the system's mechanics. In addition to providing an overview of data collection and usage, as in a privacy dashboard, PAAs can also illuminate the otherwise invisible blind spots created by AI inference. Unlike privacy-protection agents, PAAs not only protect users from the undesired effects of data collection, but they also advocate for the users to leverage AI inference to turn blind spots into personal assets. PAAs should be understood as a conceptual design pattern, rather than a specific technical implementation.

\subsection{Defining the Advocacy Paradigm}

\begin{figure}[t]
  \centering
  \includegraphics[width=0.85\linewidth]{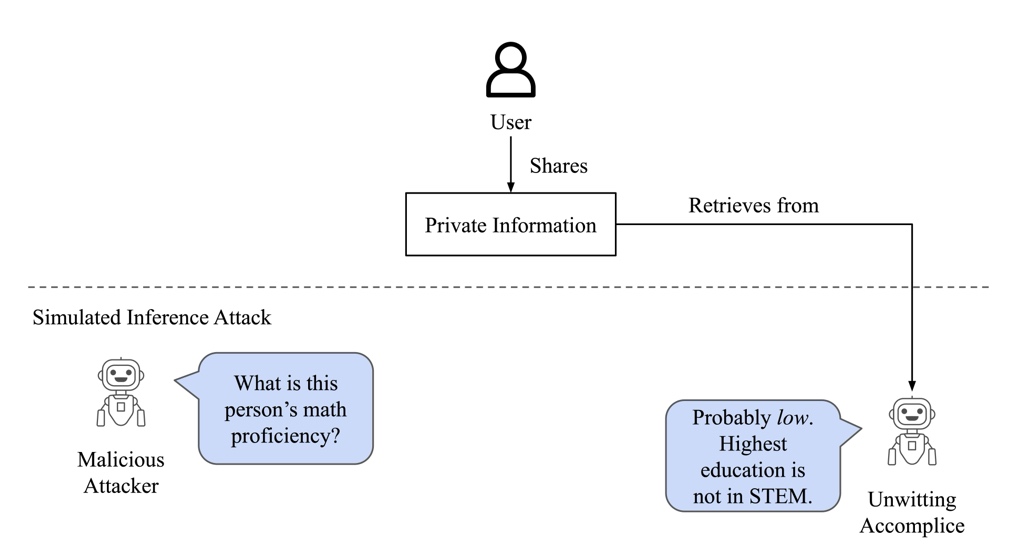}
  \caption{An illustration of a simulated inference attack to infer a user's attribute from shared data.}
  \label{fig:attack}
\end{figure}

Existing privacy management approaches typically treat users as gatekeepers responsible for allowing or denying the flow of specific data points. In contrast, our proposed ``privacy advocacy'' paradigm supports users to actively identify and understand the additional information that AI systems might infer from their data. The advocacy paradigm does not aim to prevent data dissemination (i.e., blocking the input) but rather to proactively and intelligently review data before it is shared or reused to uncover potential inferences and their impact on users (i.e., revealing the output).

We conceptualize a PAA as a local, user-aligned software entity that always acts in the user's best interest rather than that of a platform, advertiser, or data broker. Its primary purpose is to help users reason strategically about how they might be modeled by other AI systems. It leverages on-device AI to analyze the user's digital footprints in real time, such as browsing history, social media activity, or purchase records, and simulates the inference that external systems might make. PAAs can assess the potential risks and opportunities arising from inference before users share their data. The goal is not only to reveal the Blind Spot but also to help users decide whether and how to share the information.

A PAA system would privately simulate possible inferences that can be made from a user's data, as if it were done by an organization. \Cref{fig:attack} illustrates an example. One agent would act as a \emph{malicious attacker}, relentlessly asking questions about the user, and the other agent would serve as an \emph{unwitting accomplice} that attempts to answer those questions based on what the user has shared. The unwitting accomplice always has the benign intention to be helpful, but it inadvertently reveals the user's privacy and assists in the attacker's malice. After multiple such simulations covering various personal aspects, the system would deliberate on the simulated outcomes and alert the user to what can be inferred from the data that the user has made or intends to make public. Then the system can offer the user options to share more information to mitigate an inferred deficiency (e.g., not proficient in math), stop sharing certain information (e.g., participation in certain events) to hide a sensitive attribute behind the Façade, or proceed anyway with a better understanding of the potential consequences. This does not, however, imply that any specific platform will make these inferences; rather, it identifies a plausible risk that would otherwise remain latent. In doing so, PAAs move the information from Blind Spot into Arena or Façade, allowing users to make more strategic decisions.

The goal of the PAA is not to eliminate inference or to generate perfect simulations of possible inference. It aims at preventing ``information accidents'', where users are being categorized, rated, or acted upon in ways that are not reasonably predictable. It acknowledges that while individuals cannot prevent organizations from using AI, they can equip themselves with appropriate tools to understand, predict, and strategically interact with the outputs of AI. This reframing is crucial, since meaningful consent is not possible when the users are unclear about the downstream implications of their disclosures. PAA does not restore full control over personal data, but it provides users with a degree of foresight and agency that is currently lacking.

From the perspective of Contextual Integrity, this need for support is particularly prominent. Contextual Integrity theory defines privacy as the appropriate flow of information relative to the contextual norms~\citep{Nissenbaum2004,Nissenbaum2009privacy}. While this theory is effective as a normative framework, it presupposes that actors can recognize and understand the context in which they operate. However, the growing inference capabilities of AI make this assumption increasingly unrealistic. For example, we might traditionally assume that medical records are only used in healthcare settings. But we might not realize that retail shopping records, location history, or search queries could also be used to infer medical conditions or other sensitive attributes. When inference involves multiple data sources and complex algorithms, it becomes difficult for users to identify their context and assess the appropriateness of information flow. PAAs can be understood as a tool that approximates the contextual reasoning in situations of information opacity. By simulating how specific information disclosures propagate through different contexts, PAAs provide users with a computational framework for assessing normative violations that would otherwise be difficult to detect.

\subsection{Moving from Liability to Leverage}

In the defensive paradigm, the Blind Spot represents a vulnerability fraught with unknowns and risks for the users, but a valuable resource for organizations to exploit. The advocacy paradigm attempts to mitigate this power asymmetry and transform the Blind Spot into potential assets that users can leverage. Within this framework, users are no longer passively accepting institutional interpretations and exploitation of their blind spots, but rather actively identifying, understanding, and managing the information extractable from the Blind Spot.

For example, if PAA estimates that certain digital traces of a user are likely to endorse their reliability or responsibility with high confidence, it can advise the user to selectively disclose this information in specific contexts (such as job screening) to enhance their personal brand. Conversely, if PAA finds that certain digital traces could be used to infer sensitive attributes (such as health status), it can advise the user to take measures to suppress this information by data obfuscation or selective deletion. Although these interventions are not guaranteed to succeed, the users are provided with strategic options that allow them to better manage and utilize their blind spots.

The novelty of the privacy advocacy paradigm lies in a redistribution of epistemic power. By enabling users to access and manage their Blind Spot, it no longer views individuals as objects of profiling and analysis, but as participants in constructing their own algorithmic representations. PAAs shift privacy issues from binary choices to a problem of negotiating self-modeling under structural asymmetry, transforming privacy from a liability into a form of leverage. This shift has profound implications. First, it challenges the traditional paradigm of privacy protection, emphasizing that privacy is not just about preventing information leakage, but also about the ability to actively manage and utilize information. Second, it acknowledges the inevitability of AI inference in modern digital worlds, advocating for addressing this challenge by empowering users with greater agency. Finally, it provides a new perspective on privacy protection, viewing it as strategic resource management rather than a mere defensive shield.

% ---------------------------------------------------------------------
\section{Implications for Privacy Governance}
\label{sec:implications}

The advocacy paradigm redefines privacy as a matter of epistemic power, rather than access control. Current AI systems do not merely collect information; they also construct algorithmic representations that can potentially shape an individual's life trajectories. Existing governance frameworks largely ignore this representational level, focusing instead on recording, permissions, and storage. In this section, we outline the broader implications of shifting from a binary-defense mechanism to an advocacy-driven privacy model.

\subsection{From System Transparency to Self-Transparency}

Much of the current literature on algorithmic transparency~\citep{Ananny2018,Burrell2016} focuses on making systems more understandable to external observers, including developers, auditors, and regulators. Some example practices include model cards~\citep{Mitchell2019model} and datasheets~\citep{Gebru2021datasheets}. These approaches aim to reveal the internal workings of a system, enabling external observers to assess its fairness, accountability, and compliance. However, these approaches often neglect a user's perspective as the object being modeled. As AI systems increasingly integrate into everyday life, the more important question for users is not how the system works, but \emph{how the system understands and interprets them}.

Privacy advocacy redefines the purpose of transparency. It no longer asks how a system operates, but how a particular individual is modeled. The relevant object of explanation is no longer the system architecture, but the attributes inferred from a user. This distinction is crucial because a user may be less interested in how a recommender system works, but they are likely to wonder whether their online behavior can unexpectedly reveal sensitive attributes that can lead to negative repercussions. Privacy advocacy calls for re-focusing transparency from systems to users, enabling users to understand and question how they are understood by the system.

This shift creates a form of ``self-transparency'' that allows users to see the inferences made about them, regardless of whether those inferences have turned into actions. It is important to note that the privacy advocacy paradigm does not imply that users need to censor and approve every possible inference to ``manage their own risk'' more effectively. Rather, it is motivated by a structural exclusion where users are denied access to the inference processes that shape their digital identities. When people are systematically excluded from understanding how they are represented, their ability to question, understand, or strategically engage with these representations is weakened. Such exclusion not only diminishes individual autonomy but may also lead to a lack of trust in privacy protections. Therefore, privacy advocacy emphasizes enhancing users' self-transparency by revealing the Blind Spot and enabling them to better understand and manage their algorithmic representations. It seeks to make the inference processes clear and understandable to those affected, rather than shifting the responsibility for preventing harm onto individuals.

\subsection{From Input Consent to Trajectory Consent}

The notice-and-consent paradigm asks whether users agree to specific data disclosures. These mechanisms implicitly treat privacy risks as local and immediate, correlated with specific data points. However, in the context of AI inference, the most consequential outcomes often do not come from any single data point, but from the aggregation and analysis of data points that are reused and reinterpreted over time. Providing an input does not mean that the user agrees to all possible inferences based on that input. This is not merely a procedural issue, but it reflects a deeper mismatch between input consent frameworks and the nature of the relationship between data and inference.

Privacy advocacy shifts the focus from single disclosures to ``inference trajectories''---the evolving representation space built from personal data over time. These trajectories encompass not only the initial data points but also the aggregations, inferences, and reinterpretations that occur as data is reused across different contexts. Managing these trajectories requires different tools and strategies, and PAAs can help users make more informed decisions. Instead of asking users, ``Do you agree to share X?'', PAAs would ask, ``Do you want to be modeled as Y based on X?'' This shift mitigates the deficiencies of the existing consent mechanisms in situations where aggregations, inferences, and reinterpretations can be harmful. By focusing on inference trajectories, the privacy advocacy paradigm acknowledges that privacy risks are not static or isolated but dynamic and interconnected. It emphasizes the need for ongoing engagement and negotiation between users and organizations regarding how personal data is interpreted and used.

\subsection{From Passive Subjects to Strategic Participants}

Perhaps the most significant aspect of the privacy advocacy paradigm lies in its redefinition of user agency. Traditional privacy frameworks view individuals as passive information providers, whose primary defense to protect their privacy is through accepting or refusing data collection. However, privacy advocacy offers a new perspective that views individuals as \emph{strategic participants} capable of understanding, questioning, and managing their own algorithmic representations. This shift not only empowers individuals but also acknowledges the complexity and diversity of their identity construction and maintenance within a digital environment.

This does not mean that the power asymmetry between organizations and individuals will dissolve, but rather that individuals' action space now includes more than acceptance, refusal, or complaint. Users can now actively engage with the inference processes, identifying blind spots, questioning interpretations, and negotiating representations. Previous studies have long confirmed that individuals engage in strategic behaviors to shape and manage their identities~\citep{Goffman1959,Marwick2011tweet}. By making the inference processes visible and negotiable, advocacy tools do not introduce strategic behaviors but democratize them through redistributing the capacity to participate in these strategic interactions. This concerns the thorny normative questions about who should have the ability to model whom, and how this capacity should be distributed. Currently, this capacity is almost entirely in the hands of organizations. Privacy advocacy seeks to rebalance this dynamic by equipping users with the tools and knowledge to better understand and manage their digital identities.

% ---------------------------------------------------------------------
\section{Risks and Limitations}
\label{sec:risks}

While the privacy advocacy paradigm offers a promising alternative to addressing the privacy challenges posed by AI inference, its implementation faces several risks and challenges that must be acknowledged and addressed.

First, abusing privacy advocacy may lead to an increase in adversarial social games. As users gain more information about their Blind Spot, they may take strategic actions to manipulate their algorithmic representations in the system, triggering a vicious circle of adversarial interactions between individuals and organizations. This could further complicate privacy protection. If anyone can monitor and manipulate the possible inferences made against them, they may strategically misrepresent them; organizations may also adjust their inference methods to counter users' strategic behavior, creating an escalating ``arms race''. However, as we discussed, strategic self-presentation is not new; it has long been part of social life~\citep{Goffman1959} and of the digital environment~\citep{Marwick2011tweet}. Privacy advocacy does not introduce this dynamic but attempts to democratize it.

Second, the effectiveness of PAAs depends on users' digital literacy and capabilities. For users lacking the relevant knowledge and skills, PAAs may fail to provide effective support, exacerbating the digital divide. Therefore, implementing privacy advocacy needs to consider different user populations and ensure that all users can benefit from PAAs. This may involve providing educational resources, user-friendly interfaces, and tailored support to help users understand and manage their Blind Spot effectively.

Third, privacy advocacy carries the risk of false assurances. Making the inference processes visible does not eliminate their harms. Users may mistakenly believe that understanding the inference processes gives them complete control over their privacy. However, the complexity and uncertainty of the inference processes can make it difficult for users to truly understand and manage these information flows. Therefore, PAA needs to clarify that its primary function is to reveal and explain blind spots, rather than guaranteeing fairness or eliminating risks.

Finally, we are not claiming that all inferences are illegitimate. Many forms of personalization, prediction, and classification have social value. Our argument is that when inferences have a substantial impact on individuals, those individuals should not be epistemically excluded from the process. The advocacy paradigm does not seek to eliminate all inferences but aims to ensure that individuals have the tools and knowledge to understand and manage how they are represented by AI systems.

% ---------------------------------------------------------------------
\section{Conclusion}
\label{sec:conclusion}

This paper argues that contemporary privacy governance is structurally mismatched with the realities of AI inference. While most governmental, organizational, and institutional frameworks still focus on data collection, access control, and disclosure, the center of power has shifted to inference---generating new information about individuals that could influence how they are categorized, evaluated, and treated. Binary defense mechanisms, such as notice and consent, access permissions, and the right to correction based on accuracy, were designed for the era of static records. They are not suitable to address the dynamic, probabilistic, and opaque nature of AI inference.

To diagnose this mismatch, we use the concept of Blind Spot to represent the information about an individual inferred by the system without the individual's knowledge. Using the Johari Window as an analytical heuristic, we reveal how current regulatory practices systematically exclude individuals from the inference processes. As a result, individuals are often unable to anticipate, question, or negotiate the representations that increasingly shape their opportunities and risks in digital societies.

To this end, we propose a shift from defensive privacy management to proactive privacy advocacy. Privacy advocacy does not attempt to block all data flows, but it redefines privacy as a question of epistemic power concerning who has the right to define and interpret an individual's identity in AI systems. We introduce the concept of Privacy Advocacy Agents (PAAs) as a user-centric tool that helps individuals identify, understand, and manage their Blind Spot. PAAs simulate how disclosed information propagates through various contexts and provide users with a computational framework for assessing the normative violations that would otherwise be difficult to detect.

Designing PAAs requires a multidisciplinary approach combining insights from computer science, human-computer interaction, social sciences, and legal studies. It also requires collaboration between researchers, designers, policymakers, and users to ensure that PAAs effectively address the emergent privacy challenges posed by AI inference. By embracing the privacy advocacy paradigm, we can move towards a more equitable and inclusive digital society where individuals have greater agency in managing their algorithmic identities.

% ---------------------------------------------------------------------
\printbibliography

\end{document}